\documentclass[aps,pre,
reprint,
 notitlepage,
 showpacs,floatfix,nofootinbib,
 superscriptaddress
 ]{revtex4-1}

 \usepackage{subfigure}
 \usepackage{amssymb}
 \usepackage{amsfonts}
 \usepackage{amsmath}
 \usepackage{amsthm}
 \usepackage{epsfig}
 \usepackage{float}
 \usepackage[usenames,dvipsnames]{color}
 \usepackage[latin1]{inputenc} 

 \usepackage{graphics,graphicx,xcolor,subfigure}

 \usepackage[hidelinks,unicode=true]{hyperref}
 \hypersetup{colorlinks=true,
 	linkcolor=blue,
 	urlcolor=blue,
 	citecolor=blue,
 	pdfhighlight=/N
 }
 
 \usepackage{comment}
 \usepackage[normalem]{ulem}
 \usepackage{microtype}

\newcommand{\Ninf}{N_\mathrm{inf}}

\newcommand{\NSI}{N_\mathrm{SI}}

\begin{document}

\title[Accuracy of discrete- and continuous-time mean-field theories]{Accuracy of discrete- and continuous-time mean-field theories for epidemic processes on complex networks}

\author{Diogo H. Silva}
\affiliation{Instituto de Ci\^{e}ncias Matem\'{a}ticas e de Computa\c{c}\~{a}o, Universidade de S\~{a}o Paulo, S\~{a}o Carlos, SP 13566-590, Brazil}

\author{Francisco A. Rodrigues}
\affiliation{Instituto de Ci\^{e}ncias Matem\'{a}ticas e de Computa\c{c}\~{a}o, Universidade de S\~{a}o Paulo, S\~{a}o Carlos, SP 13566-590, Brazil}

\author{Silvio C. Ferreira}
\affiliation{Departamento de F\'{i}sica, Universidade Federal de Vi\c{c}osa, 36570-900 Vi\c{c}osa, Minas Gerais, Brazil}
\affiliation{National Institute of Science and Technology for Complex Systems, 22290-180, Rio de Janeiro, Brazil}

\begin{abstract}
Discrete- and continuous-time approaches are frequently used to model the role of heterogeneity on dynamical interacting agents on the top of complex networks. While, on the one hand, one does not expect drastic differences between these approaches, and the choice is usually based on one's expertise or methodological convenience, on the other hand, a detailed analysis of the differences is necessary to guide the proper choice of one or another approach. We tackle this problem, by comparing both discrete- and continuous-time mean-field theories for the susceptible-infected-susceptible (SIS) epidemic model on random networks with power-law degree distributions. We compare the discrete epidemic link equations (ELE) and continuous pair quenched mean-field (PQMF) theories with the corresponding stochastic simulations, both theories that reckon pairwise interactions explicitly. We show that ELE converges to the PQMF theory when the time step goes to zero. We performed an epidemic localization analysis considering the inverse participation ratio (IPR). Both theories present the same localization dependence on network degree exponent $\gamma$: for $\gamma<5/2$ the epidemics are localized on the maximum k-core of networks with a vanishing IPR in the infinite-size limit while for $\gamma>5/2$, the localization happens on hubs that do form a densely connected set and leads to a finite value of IPR. However, the IPR and epidemic threshold of ELE depend on the time-step discretization such that a larger time-step leads to more localized epidemics. A remarkable difference between discrete and continuous time approaches is revealed in the epidemic prevalence near the epidemic threshold, in which the discrete-time stochastic simulations indicate a mean-field critical exponent $\theta=1$ instead of the value $\theta=1/(3-\gamma)$ obtained rigorously and verified numerically for the continuous-time SIS on the same networks.\\


\end{abstract}

\maketitle


\section{Introduction}
The broad diversity of spreading processes~\cite{Romualdo2015,DeArruda2018} and the role that they have assumed in our contemporary society call for the development of theoretical tools~\cite{Wang2017} to predict outbreaks and ways to mitigate their consequences~\cite{Desai2019}. For example, the recent epidemic outbreaks of Zika virus~\cite{Angel2017}, Ebola~\cite{Gomes2014}, and  COVID-19 pandemics~\cite{Arenas2020,Guilherme2020} enhanced the development of compartmental models~\cite{Keeling2011,Romualdo2015} for epidemic processes on populations.  

The heterogeneity of contacts among individuals influences the spreading process~\cite{Romualdo2015,DeArruda2018} and these connection patterns can be tackled using complex networks theory~\cite{Barrat2008,Newman2010}. Theoretical frameworks for epidemic models running on the top of complex networks allow one to investigate the coupling between the dynamic and structural features of the system.  In continuous-time approaches, different mean-field theories are widely applied to address these features. The whole network structure is considered in the quenched mean-field (QMF)~\cite{DeArruda2018,Wang2017} theory that neglects dynamical correlations of the epidemic states of connected vertices. This theory is improved by considering correlations in a pairwise level in the pair quenched mean-field (PQMF)~\cite{Mata2013,Diogo2019} theory. These theories have been applied to investigate the limit of low epidemic prevalence (average fraction of infected nodes)~\cite{VanMieghem2012_a, VanMieghem2012_b, Diogo2019, Goltsev2012,Mata2013}, while the PQMF theory also presents an excellent accuracy in the regime of high prevalence~\cite{Diogo2020}. 

Some aspects of actual epidemic processes, such as the daily update of reported cases by surveillance systems~\cite{Desai2019} and computational facilities~\cite{LI2010, Gomez2010, Gomez2011, Clara2013, Arenas2020, Matamalas2018}, lead to the wide use of discrete-time versions of epidemic dynamics. Note that a straightforward discretization of continuous-time theoretical approaches presents limitations~\cite{Gomez2011} as, for example, the epidemic prevalence in the supercritical regime converging to nontrivial fixed points~\cite{Jun2013} $-$ the more striking for longer time steps~\cite{Wang2019}. Pitfalls of the discrete-time framework can be fixed using the nonlinear dynamical system (NDLS) approach~\cite{Wang2003,Chakrabarti2008}, which explores the representation of the dynamics as Markov chains. This approach was generalized in the so-called microscopic Markov-chain approach (MMCA)~\cite{Gomez2010}, which was adapted and applied to different problems, such as propagation of information and epidemics on multiplex networks~\cite{Clara2013, Paulo2019,Wang2021}, metapopulation~\cite{Gomez2018, Cota2021} modeling of spatiotemporal epidemic spreading~\cite{Arenas2020, Soriano2022}. In particular, a discrete-time counterpart of the PQMF theory~\cite{Mata2013},  adding dynamical correlations to the MMCA, called epidemic linking equations (ELE)~\cite{Matamalas2018}, was applied to investigate optimal immunization strategies in complex networks.

Epidemic or, more generally, dynamical processes on networks, are featured by localization phenomena where the activity is highly concentrated on a finite~\cite{Goltsev2012,Diogo2021} or sub-extensive~\cite{Pastor-Satorras2018,Diogo2021} part of the network. Localization in epidemic processes was formerly associated with the leading eigenvector of the adjacency matrix that rules the local epidemic prevalence of the QMF theory~\cite{Goltsev2012}. The idea was extended to PQMF~\cite{Diogo2019} and, later, to generic epidemic processes on networks~\cite{Diogo2021}, including stochastic simulations. The accuracy of mean-field approaches is correlated with the level of localization of the activity near the epidemic threshold in each theory~\cite{Diogo2019,Silva2022}. The more localized the outcome of a mean-field theory, the less accurate its prediction of the epidemic threshold~\cite{Diogo2019}. 

In this paper, we develop a numerical study of the accuracy of discrete-time theoretical approaches represented by  MMCA and ELE in comparison with statistically exact stochastic simulations for both continuous- and discrete-time approaches.  We focus on determining the epidemic threshold and prevalence of the susceptible-infected-susceptible model (SIS)~\cite{Romualdo2015} in uncorrelated networks with power-law degree distributions, where an infected individual heals spontaneously with rate $\mu$ and transmits the disease to its susceptible neighbors with rate $\beta$ (continuous-time version) per contact.  This simple model presents far from trivial features. For example, the epidemic threshold is null in the thermodynamic limit, where the system size goes to infinity, in the case of power-law degree distribution regardless of the degree exponent $\gamma$~\cite{Mountford2013,Chatterjee2009}. In the discrete-time SIS model, the healing and infection probabilities are given by $g=\mu \Delta t$ and $r=\beta \Delta t$, respectively. When these probabilities assume large values,  the discrete-time SIS deviates substantially from its continuous-time version~\cite{Fennell2016}. Indeed, in the limit of high infection probability and prevalence, Chang and Cai~\cite{Chang2021} claimed that there is no linear mapping from discrete- to continuous-time dynamics while Zhang et al.~\cite{Zhang2022} proposed a nonlinear relation between rates and probabilities to circumvent this drawback. 

Different theoretical approaches are compared with either optimized Gillespie (OGA)~\cite{Cota2017} or synchronous update (SUA)~\cite{Fennell2016}  algorithms for continuous and discrete time simulations, respectively. We report that, as in the continuous-time limit, the introduction of dynamical correlation in ELE improves the accuracy of MMCA approach to describe the epidemic threshold and epidemic prevalence,
and the highest accuracy happens in the continuous-time limit when ELE converges to PQMF theory. We also performed an epidemic localization analysis~\cite{Diogo2021} and a non-perturbative (high prevalence regime) numerical integration, showing that ELE presents the same characteristics of the continuous time limit obtained with PQMF theory~\cite{Diogo2019}: localization on the maximum k-core~\cite{Castellano2012} for $\gamma<5/2$ and on a finite set of nodes (hubs) for $\gamma>5/2$ near to the epidemic threshold (low-prevalence regime) concomitant with a very good accuracy in the high-prevalence regime. Finally,  discrete-time (no small time steps) stochastic simulations are consistent a mean-field critical exponent $\theta=1$ associated to vanishing of the epidemic prevalence near to the epidemic threshold instead of the value $\theta=1/(3-\gamma)$ obtained rigorously~\cite{Chatterjee2009} and verified numerically~\cite{Diogo2019} in the continuous-time simulations of SIS model on the same set of power-law networks.

The remainder of this paper is organized as follow: In sec~\ref{sec:theory}, we present the continuous- and discrete-time theories and demonstrate that ELE (MMCA) converges to PQMF (QMF)  in the limit  $\delta t \rightarrow 0$. Section~\ref{sec:result} starts with a discussion about the accuracy of continuous and discrete theories in predicting the epidemic threshold. This analysis is extended to localization and epidemic prevalence. Finally, in sec.~\ref{sec:conclu}, we draw our conclusions and prospects.

\section{Analytical and simulation approaches}
\label{sec:theory}

\subsection{Continuous-time mean-field theories}

In the QMF theory, the whole network structure is considered by the introduction of the adjacency matrix, defined as $A_{ij}=1$ if nodes $i$ and $j$ are connected and $A_{ij}=0$ otherwise, while dynamical correlations are neglected assuming that that probability $\rho_i$ that a node $i$ is infected does not depend on the states of its contacts. Thus, the dynamical equation for evolution of $\rho_i$ is~\cite{Romualdo2015,DeArruda2018,Wang2017}
\begin{equation}
	\frac{d\rho_{i}}{dt}=-\mu\rho_{i}+ \beta(1-\rho_{i})\sum_{i=1}^{N}A_{ij}\rho_{j},
	\label{eq:QMF1}
\end{equation}
where  the first term on the right-hand side corresponds to spontaneous healing  while the second one is the infection. If dynamical correlations are taken into account in a pair level in the named PQMF theory~\cite{Mata2013,Diogo2019}, dynamical equation becomes
\begin{equation}
	\frac{d\rho_{i}}{dt}=-\mu\rho_{i}+ \beta\sum_{i=1}A_{ij}\phi_{ij}, 
	\label{eq:PQMF1}
\end{equation}
in which $\phi_{ij}=[S_{i},I_{j}]$ is the probability that nodes $i$ and $j$ are in the states susceptible and infected, respectively. Note that, $\phi_{ij}\approx (1-\rho_{i})\rho_{j}$  leads to the QMF theory, Eq.~\eqref{eq:QMF1}. Following Ref.~\cite{Mata2013}, the evolution of $\phi_{ij}$ is given by 
%
\begin{eqnarray}
	\frac{d\phi_{ij}}{dt}=-(2\mu+\beta)\phi_{ij}&+&\mu \rho_{j}+\beta \sum_{l=1}\frac{\omega_{ij}\phi_{jl} }{(1-\rho_{j})}(A_{jl}-\delta_{il}) \nonumber\\&-& \beta \sum_{l=1}\frac{\phi_{ij}\phi_{il} }{(1-\rho_{i})}(A_{il}-\delta_{lj}),
	\label{eq:PQMF2}
\end{eqnarray}
in which $\omega_{ij}=[S_i,S_j] = 1-\rho_{i}-\phi_{ij}$. 
Close enough to the epidemic threshold, the QMF and PQMF theories can be described by the spectral properties of the adjacency matrix in the former~\cite{Goltsev2012,VanMieghem2012_a} and a weighted adjacency matrix in latter~\cite{Diogo2019}.

\subsection{Discrete-time mean-field theories}

The general form for the temporal evolution of the probability that a node $i$ is infected, i.e., $\rho_{i}$, in the MMCA~\cite{Gomez2010} assumes the form 
\begin{eqnarray}
	\rho_{i}(t+\Delta t)= (1-\mu\Delta t)\rho_{i}(t)&+&[1-q_{i}(t)][1-\rho_{i}(t)] \nonumber\\
	&+& \mu\Delta t[1-q_{i}(t)]\rho_{i}(t), 
	\label{eq:MMCA1}
\end{eqnarray}

in which $q_{i}(t)$ is the probability that node $i$  was not infected in the corresponding time  step
\begin{equation}
	q_{i}(t)=\prod_{j=1}^{N}[1- \beta\Delta t A_{ij}\rho_{j}(t)].
	\label{eq:MMCA2}
\end{equation} 
In Eq.\eqref{eq:MMCA1}, the first term stands for the probability that node $i$ remains infected while in the second term is the probability that a susceptible node $i$ is infected by any of its contacts. Finally, the last term reckons the probability that an infected node recovers and is reinfected by one of its neighbors during the time step $\Delta t$. In the present work, we disregard the reinfection term,  of order $(\Delta t)^2$, in MMCA that does not change either the epidemic threshold nor the continuous-time limit of the theory. Thus, we can compare both MMCA and ELE with the same synchronous discrete-time simulations (see subsection~\ref{subsec:simu}). So, the MMCA equation investigated is~\cite{Gomez2011a}
\begin{align}
	\rho_{i}(t+\Delta t)= (1-\mu\Delta t)\rho_{i}(t)+[1-q_{i}(t)][1-\rho_{i}(t)].
	\label{eq:MMCA3}
\end{align} 
Note that an oscillating (period 2) epidemic prevalence is observed in both MMCA~\cite{Gomez2011a} and ELE~\cite{Matamalas2018} theories without reinfections and the steady-state prevalence is calculated as the arithmetic average of the values.

In the pairwise ELE approach~\cite{Matamalas2018}, the system is described in terms of the joint probabilities  $\Phi_{ij}=\phi_{ij}=[S_{i}, I_{j}], \Theta^{\text{S}}_{ij} =\omega_{ij}=[S_{i}, S_{j}]$ and $\psi_{ij}=\Theta^{\text{I}}_{ij}=[I_{i}, I_{j}]$ given the states of connected pairs of node $i$ and $j$. In particular, the evolution of $\phi_{ij}$ is derived in Ref.~\cite{Matamalas2018} and given by
\begin{widetext}
\begin{equation}
	\phi_{ij}(t+\Delta t)  = \omega_{ij}(t)q_{ij}(t)[1-q_{ji}(t)]
	+\mu\Delta t(1-\mu\Delta t)\psi_{ij}(t)+\mu\Delta t[1-(1-\beta \Delta t)q_{ji}(t)]\phi_{ji}(t)
	+(1-\mu\Delta t)(1-\beta \Delta t)q_{ij}(t)\phi_{ij}(t)
	\label{eq:ELE1}
\end{equation}
\end{widetext}
where,
\begin{eqnarray}
	q_{ij}(t)=\prod_{\substack{l=1 \\ l\neq j} }^{N} \left(1-\beta\Delta t A_{li} \frac{\phi_{il}(t)}{1-\rho_{i}(t)}\right)
	\label{eq:ELE2}
\end{eqnarray}
is the probability that a susceptible node $i$ is not infected by any of its neighbors (excluding node $j$) in the step $t$. Similar equations can be built for other joint probabilities; see Ref.~\cite{Matamalas2018}. The epidemic prevalence in the whole system is expressed as~\cite{Matamalas2018}
\begin{equation}
	\rho=\frac{1}{N}\sum_{i=1}^{N}\frac{1}{k_{i}}\sum_{j=1}^{N} A_{ji}(\phi_{ji}+\omega_{ij})
	\label{eq:ELE3}
\end{equation}
where $k_{i}$ is the degree of node $i$.

The effects of dynamical correlations in ELE theory become  clearer for the variable $\rho_{i}=\psi_{ij}+\phi_{ji}$ that  evolves as 
\begin{eqnarray}
	\rho_{i}(t+\Delta t)=(1-\mu\Delta t)\rho_{i}(t)&+&[1-q_{ij}(t)][1-\rho_{i}(t)] \nonumber\\
	&+&\beta\Delta t q_{ij}(t)\phi_{ij}(t).
	\label{eq:ELE4}
\end{eqnarray}
The first and second terms have the similar form and interpretation of Eq.\eqref{eq:MMCA1}. The last term stands for the probability that nodes $i$ and $j$ are susceptible and infected, respectively, and node $i$ is infected by node $j$.

The continuous- and discrete-time mean-field theories are equivalent in the regime of $\Delta t\rightarrow 0$. We now show the equivalence between ELE and PQMF while one  can easily demonstrate the equivalence between MMCA and QMF following similar steps. To leading order in $\Delta t$,  Eq.~\eqref{eq:ELE2} becomes
\begin{equation}
	q_{ij}(t)=1-\beta\Delta t \sum_{r=1}^{N}\frac{\phi_{ir}(t)}{1-\rho_{i}(t)}(A_{ri}-\delta_{rj})+\mathcal{O}(\beta\Delta t),
	\label{eq:ELE5}
\end{equation}
where $\mathcal{O}(x)$ represent terms negligible if compared with $x$. 
Substituting~Eq.~\eqref{eq:ELE5} in Eqs.~\eqref{eq:ELE1} and \eqref{eq:ELE4} leads to
\begin{equation}
	\frac{\rho_{i}(t+\Delta t)-\rho_{i}(t)}{\Delta t}=-\mu  \rho_{i}(t) +\beta\sum_{r}A_{ri}\phi_{ir} +O(\beta\Delta t) 
	\label{eq:ELE6}
\end{equation}
and 
\begin{widetext}
\begin{equation}
	\frac{\phi_{ij}(t+\Delta t)-\phi_{ij}(t)}{\Delta t} = 
	\psi(t)-(\beta+\mu) \phi_{ij}(t) +\beta\sum_{l=1}^{N}\frac{\omega_{ij}(t)\phi_{jr}(t)}{1-\rho_{j}(t)}(A_{rj}-\delta_{ri}) -\beta \sum_{l=1}^{N}\frac{\phi_{ij}(t)\phi_{ir}(t)}{1-\rho_{i}(t)} (A_{ri}-\delta_{rj})+O(\beta\Delta t)
	\label{eq:ELE7}
\end{equation}
\end{widetext}
Finally, considering the relations $\rho_{j}=\psi_{ij}+\phi_{ij}$ and $1-\rho_{i}=\omega_{ij}+\phi_{ij}$, in the limit of $\Delta t\rightarrow 0$, Eqs.~\eqref{eq:ELE6} and \eqref{eq:ELE7} converge to Eqs.~\eqref{eq:PQMF1} and \eqref{eq:PQMF2}. 

\subsection{Stochastic simulations}
\label{subsec:simu}

QMF and PQMF approaches are compared with  continuous-time stochastic simulations  implemented using the optimized Gillespie algorithm (OGA)~\cite{Cota2017}. In this algorithm, the probabilities of healing and infection are determined by the number of infected nodes $\Ninf$ and the total number of edges emanating from them $\NSI$.  In each time step, with probability
\begin{equation}
	q=\frac{\mu \Ninf}{\mu\Ninf+\beta\NSI},
	\label{eq:GA1}
\end{equation}
an infected node is chosen at random and healed. With probability $1-q$ an infected node $i$ is chosen with probability proportional to its degree $k_i$. If the randomly chosen neighbor $j$ is susceptible, it becomes infected; otherwise no change os state is implemented. The time is incremented by 
\begin{equation}
	\delta t  = \frac{-\ln u}{\mu\Ninf+\beta\NSI}
\end{equation}
where $u$ is a pseudo-random number  uniformly distributed in the interval $(0,1)$.

The discrete-time simulations are performed using the synchronous updating  algorithm (SUA)~\cite{Cai2014,Chang2021,Fennell2016}, in which all nodes of the network have their states simultaneously updated in a discrete time step, $\Delta t$. For the SIS dynamics, each susceptible node becomes infected  with probability $1-(1-\beta \Delta t)^{n_{\text{inf}}}$, in which $n_{\text{inf}}$ is the number of infected contacts while an infected node becomes susceptible with probability $\mu \Delta t$. Finally, the time is incremented by $\Delta t$. 

In both, continuous and discrete simulations, when the system falls into the absorbing state, the dynamics returns to a previously visited active configuration using the standard quasi-stationary (QS) method~\cite{Oliveira2005,Sander2016}. In this method, a list of $M$ configurations is built and constantly updated by replacing a randomly chosen configuration by the the current one with a probability $P_{\text{rep}}$ by unit of time. In both cases, we adopted $M=50$ and $P_{\text{rep}}=0.01$. The the QS averages were performed considering the averaging time varying from $t_\text{av}=10^5$ to $10^6$ time units after a relaxation time  $t_\text{rlx}=10^5$ time units, the longer averaging and relaxation times for the lower densities where fluctuations are more relevant. To compare the mean-field theories  with QS simulations,  we estimate the epidemic threshold  using the peak of dynamical susceptibility defined $\chi=N(\langle \rho^{2}\rangle-\langle \rho\rangle^{2})/\langle \rho\rangle$~\cite{Ferreira2012}. 

\section{Results}

\label{sec:result} 

We investigate both mean-field theories and stochastic simulations on synthetic networks with power-law degree distributions for different degree exponents generated by the uncorrelated configuration model (UCM)~\cite{Catanzaro2005}.  This model has upper cutoff $k_\text{c}\lesssim \sqrt{N}$ that guarantees the absence of degree correlations for very large networks. {In principle, dynamics very localized around hubs can produce multiple peaks in $\chi(\lambda)$ curves~\cite{Ferreira2012}. Here, we used networks with cutoffs in the degree distribution that avoid the multiplicity of peaks; see Ref.~\cite{Diogo2019} for more details.}

\subsection{Epidemic threshold}

The MMCA and QMF predict the same epidemic threshold for the SIS dynamics, given by~\cite{Romualdo2015,Gomez2010}
\begin{equation}
	\lambda_{\text{c}}=\frac{\beta}{\mu}=\frac{1}{\Lambda_{1}}
	\label{eq:threshold1}
\end{equation}
in which, $\Lambda_{1}$ is the largest eigenvalue of the adjacency matrix. The epidemic threshold of the ELE approach is given by~\cite{Matamalas2018}
\begin{equation}
	\lambda_{\text{c}}=\frac{\beta}{\mu}=\frac{1}{\Omega_{1}}.
	\label{eq:threshold2}
\end{equation}
where $\Omega_{1}$ the largest eigenvalue  of the matrix $B_{ij}$ given by
\begin{equation}
	B_{ij}=\left(1-\Upsilon\right)A_{ij} -\Upsilon k_{i}\delta_{ij}
	\label{eq:threshold3}
\end{equation}
in which, $\delta_{ij}$ is the Kronecker delta function and  
\begin{equation}
	\Upsilon=\frac{\beta(1-g)}{\mu(2-g)+2\beta(1-g)}.
	\label{eq:threshold4}
\end{equation}

\begin{figure}[!h]
	\centering
	\includegraphics[width=0.9\linewidth]{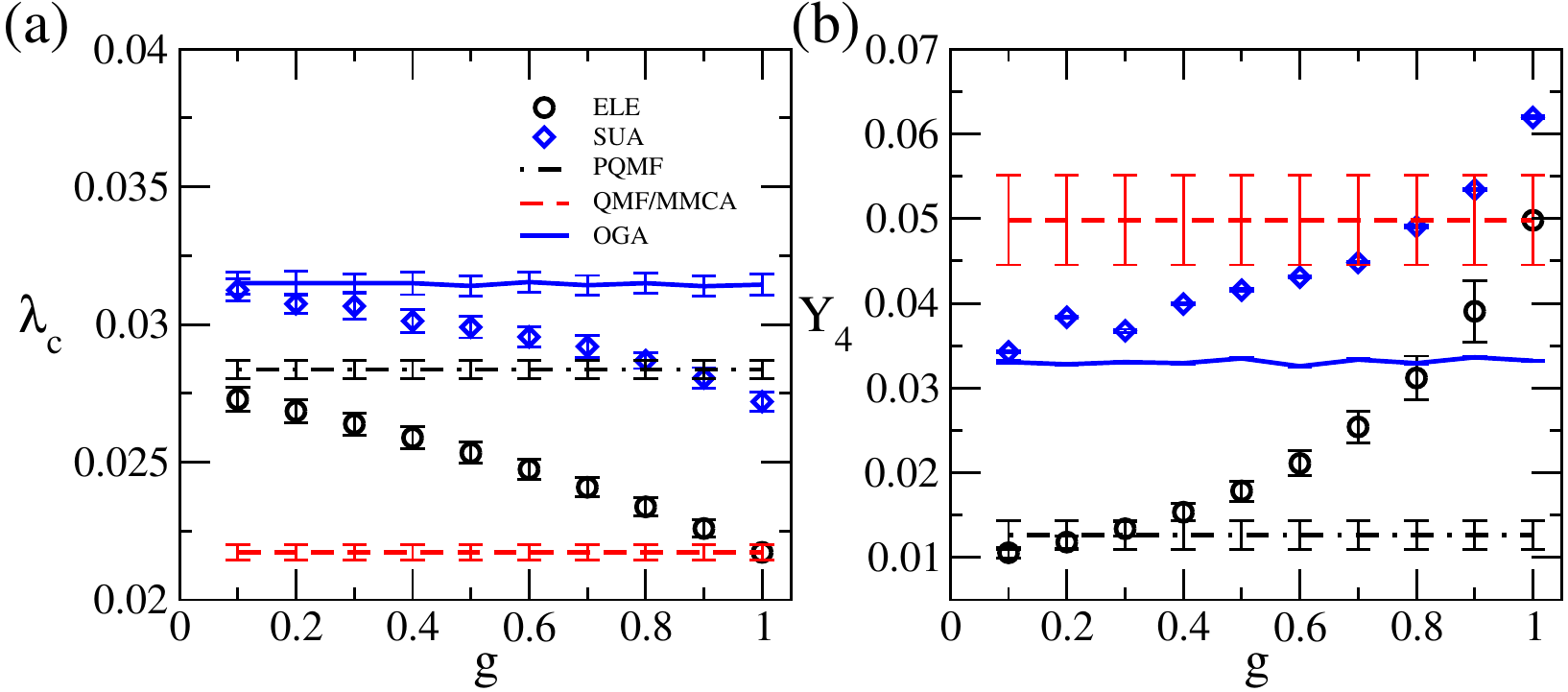}	
	\caption{Comparison of stochastic simulations and mean-field theories for SIS dynamics on uncorrelated PL networks with $\gamma=2.8$, $N=10^{6}$ and $k_{c}=2\sqrt{N}$. (a) Epidemic threshold and (b) IPR are shown as functions of $g=\mu \Delta t$ with $\mu=1$. Symbols and lines represent discrete- and continuous-time versions of stochastic simulations and theoretical frameworks, respectively. In (b), the IPR of both OGA and SUA stochastic simulations correspond to NAV multiplied by 10, for the sake of visibility. Averages were computed over 10 network-independent realizations.}
	\label{fig:limiar1}
\end{figure}

For the PQMF theory, the epidemic threshold is obtained when the largest eigenvalue of the matrix~\cite{Mata2013}
\begin{equation}
	L_{ij}=-\left(\mu+\frac{\beta^{2}k_{i}}{2\mu+2\beta} \right)\delta_{ij}+ \frac{\beta (2\mu+\beta)}{2\mu+2\beta}A_{ij},
	\label{eq:threshold5}
\end{equation}
is null. When $g=\mu \Delta t\rightarrow 0$ we have $\Upsilon= \beta/(\mu+2\beta)$ and
\begin{equation}
	L_{ij}=\mu\delta_{ij}-\beta B_{ij},
\end{equation}
where we confirm that the epidemic threshold of ELE  converges to the PQMF  theory. Moreover, for $g\rightarrow 1$ we have that $B_{ij}\rightarrow A_{ij}$  and ELE threshold goes to the same value of the QMF theory. Figure~\ref{fig:limiar1}(a) presents the ELE and MMCA epidemic thresholds as functions of $g$ for fixed $\mu=1$ as well as the results of PQMF theory with $\mu=1$, where one can see the ELE's threshold decreases monotonically from the PQMF to the QMF theory result as $g$ increases from 0 to 1.

Figure~\ref{fig:limiar1}(a) compares simulations and mean-field theories on a power-law network with $\gamma=2.8$, $N=10^{6}$ nodes and $k_\text{c}=2\sqrt{N}$. Discrete- and continuous-time simulations depart from each other as the time increment increases. The discrete case threshold decays with $g=\mu\Delta t$, showing a nonphysical dependence on the method as already discussed in Ref.~\cite{Fennell2016}. Observe that ELE reproduces qualitatively the decays of the threshold observed on discrete-time stochastic simulations (SUA), which does not happen with MMCA that provides a constant threshold. The ELE approximation deviates more from the discrete-time simulations as the time step increases; the highest precision happens in the limit $g\ll 1$ when it converges to the PQMF theory. These results hold for other values of $\gamma$ (data not shown), presenting a greater accuracy whether $\gamma<2.5$.

\subsection{Localization}

To determine the localization pattern of both simulations and mean-field theories, we use the normalized activity vector (NAV) defined as~\cite{Diogo2021}
\begin{equation}
	\varphi_{i}=\frac{\rho_{i}}{\sqrt{\sum_{j=1}^{N} \rho_{j}}},
	\label{eq:local1}
\end{equation}
where the local prevalence $\{\rho_i\}$ is computed at the epidemic threshold for different approaches (continuous, discrete, mean-field, and simulations). The localization of the NAV is quantified using the inverse partition ratio (IPR)~\cite{Goltsev2012}  defined as 
\begin{equation}
	Y_{4}(\boldsymbol{\varphi})=\sum_{i=1}^{N}\varphi_{i}^{4}.
	\label{eq:local3}
\end{equation} 
The IPR converges to a finite value in the infinite-size limit if the localization happens on a finite set of nodes, and scales as $Y_4\sim N^{-1}$ for a delocalized state involving an extensive component of the network.  For MMCA~\cite{Gomez2010} and QMF mean-field~\cite{Goltsev2012} approaches the local prevalence at the epidemic threshold is asymptotically proportional to the principal eigenvector (PVE), corresponding to the largest eigenvalue of the adjacency matrix. In the PQMF theory, the epidemic prevalence is proportional to the PVE of the Jacobian matrix $L_{ij}$, Eq.~\ref{eq:threshold5},  evaluated at $\lambda=\lambda_{\text{c}}$ as shown in Ref.~\cite{Diogo2019}. These relations are easily obtained within each theoretical approach as illustrated for ELE in the sequence of the paper. 

Very close to the epidemic threshold, the local prevalence is given by (See Eq. (17) of Ref.~\cite{Matamalas2018})
\begin{equation}
	\rho_{i} = \frac{\beta}{\mu}\sum_{j}B_{ij}\rho_j.
\end{equation}
Let $\Omega_1\ge\Omega_2\ge,\ldots, \Omega_N$ be the eigenvalues corresponding to the  eigenvectors $\mathbf{b}^{(1)},\mathbf{b}^{(2)},\ldots,\mathbf{b}^{(N)}$ of $B_{ij}$. Expanding $\rho_j = \sum_{l} c^{(l)}b_i^{(l)}$ in the basis $\{\mathbf{b}^{(l)}\}$, one obtains
\begin{eqnarray}
	\rho_i=\frac{\beta}{\mu} \sum_{l}c^{(l)}\sum_iB_{ij}b_j^{(l)} = 
	\frac{\beta}{\mu}\sum_l \Omega_l c^{(l)} b_i^{(l)} \nonumber\\ 
	\rho_i \approx \frac{\beta}{\mu}\Omega_1  b^{(1)}_i\left[1+\mathcal{O}\left( \frac{\Omega_2}{\Omega_{1}
	}\right)\right].
\end{eqnarray}
If $B_{ij}$ has a spectral gap $\Omega_{1}\gg\Omega_{2}$, that is usually the case investigated in the present paper, we find $\rho_i\sim b_i^{(1)}$. For simulations, we consider the QS local prevalence evaluated at the epidemic threshold~\cite{Diogo2021}. Figure~\ref{fig:limiar1}(b) shows the  IPR as a function of $g$ for different approaches. Corresponding values for the continuous time are shown for the sake of comparison. 
\begin{figure}[ht]
	\centering
	\includegraphics[width=0.9\linewidth]{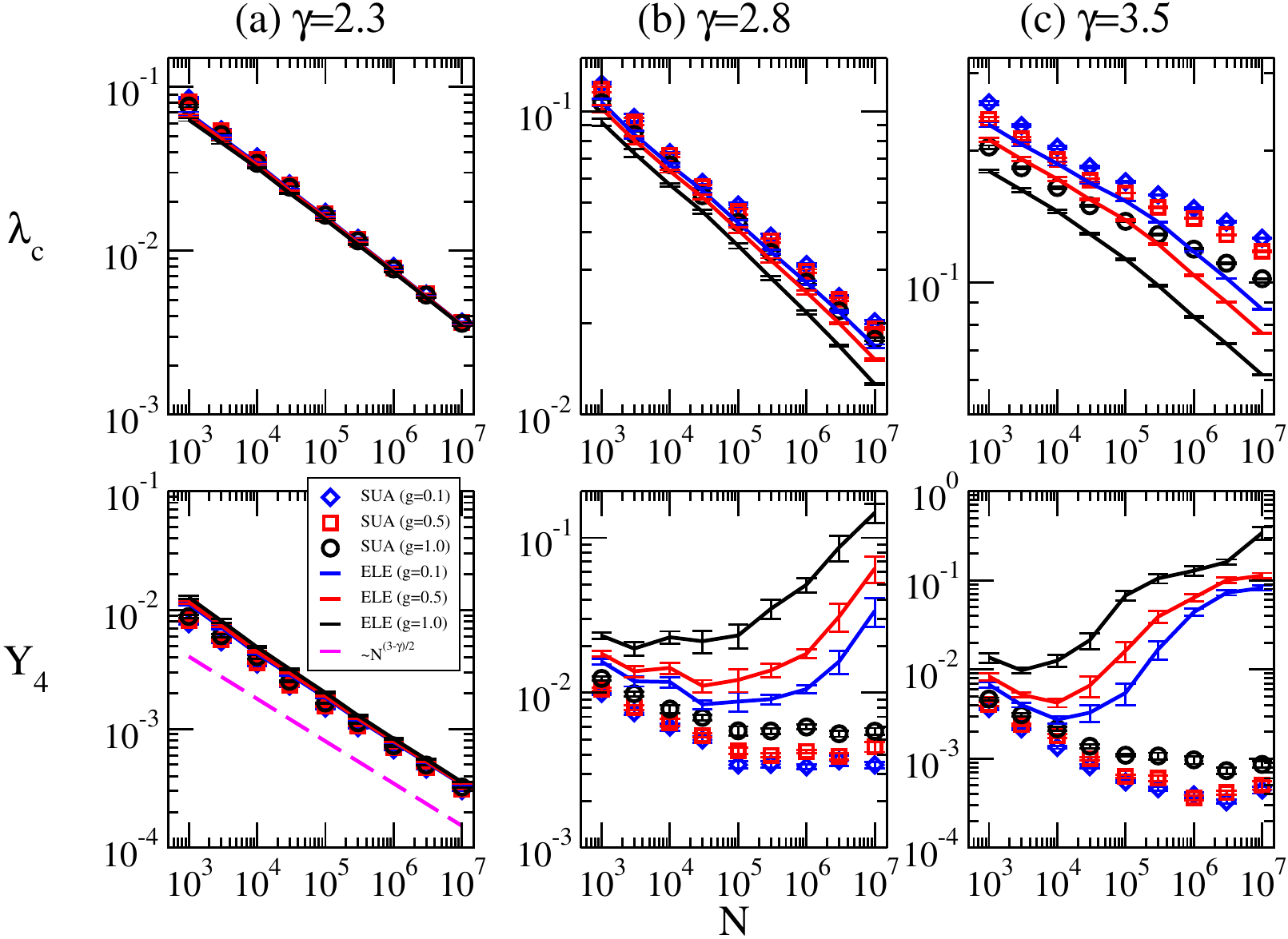}
	\caption{Finite-size analysis of epidemic threshold (top) and IPR (bottom) for ELE and discrete time simulations (SUA). Power-law networks with degree exponents (a,d) $\gamma=2.3$, (b,e) $\gamma=2.8$, and (c,f) $\gamma=3.5$ and three values of $g=0.1$, $0.5$, and $1$ are presented.  The blue dashed line represents $Y^{4}\sim N^{-(3-\gamma)/2}$. We adopted $k_{c}=2\sqrt{N}$ and $k_{c}\sim N^{1/\gamma}$ for $\gamma<3$ and $\gamma>3$, respectively.}
	\label{fig:ipr}
\end{figure} 
While MMCA has a localization that does not depend on the time-step size, the epidemics become more localized as the time step increases for both ELE and SUA simulations, being minimal in the continuous time-limit $g\rightarrow 0$ and maximal when $g\rightarrow 1$. Note that the PVE of $B_{ij}$ converges to that of $L_{ij}$ when $g\rightarrow 0$ and to that of $A_{ij}$ when $g\rightarrow 1$; see Eq.~\eqref{eq:threshold3}. Therefore, we have seen that discrete approaches, both simulation and ELE,  suffer from stronger localization effects when compared with their continuous-time counterparts.

A detailed finite-size analysis of the epidemic threshold and localization in ELE and discrete-time simulations are presented in Fig.~\ref{fig:ipr}. For $\gamma=2.3$ shown Figs.~\ref{fig:ipr} (a) and (d), we have an agreement between simulations and theory and an independence with the time-step size. This behavior is indeed expected for $\gamma=2.3$ since the ELE epidemic threshold and PVE interpolate between QMF and PQMF theories as $g$ varies in the range $(0,1)$ while the continuous-time mean-field theories match each other for $\gamma<5/2$~\cite{Mata2013,Diogo2019}. The IPR is consistent with a sub-extensive localization characterized by sublinear scaling $Y_{4}\sim  N^{-(3-\gamma)/2}$ what depicts an activity localized in the maximum $k$-core~\cite{Diogo2021,Satorras2016}, which is densely connected subgraph accessed by a $k$-decomposition~\cite{Dorogovtev2006}. 
For $\gamma>5/2$, however, the localization increases with size in ELE and saturates for stochastic simulations. Concomitantly, we observed a worse performance of the ELE to predict the epidemic threshold as the network size increases, the worse for larger $\gamma$ where localization is stronger. We also see that the finite time-steps raise the localization effects, reducing ELE accuracy as also shown in Fig.~\ref{fig:limiar1}.  This drawback should be taken into account in the choice of the theoretical approach to model the epidemics.

\subsection{Epidemic prevalence}

\begin{figure}[h]
	\centering
	\includegraphics[width=0.6\linewidth]{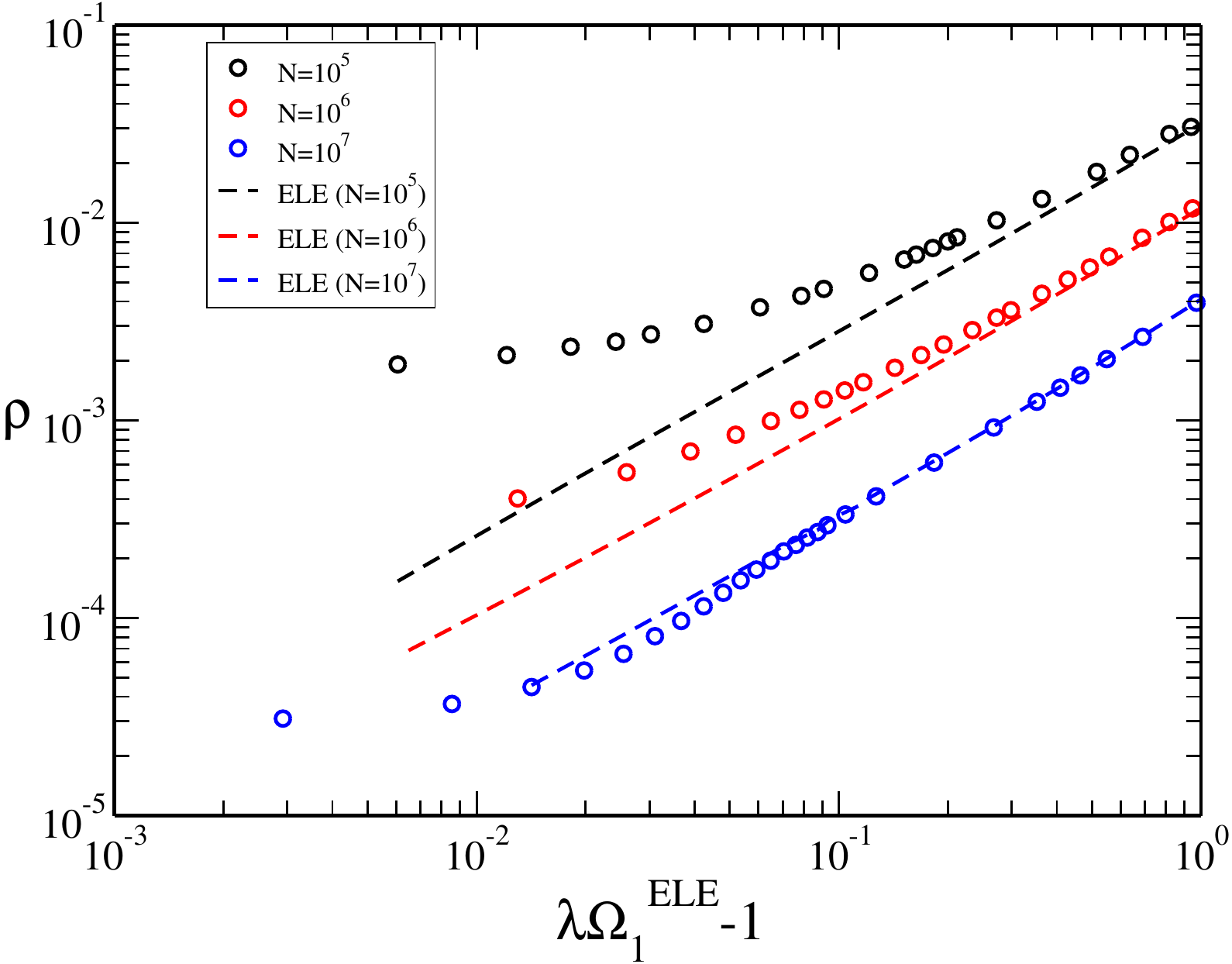}
	\caption{Comparison of epidemic prevalence for SUA simulations (symbols) and ELE (dashed lines) mean-field theory with $g=\mu\Delta t =0.5$. We consider power-law networks with degree exponent $\gamma=2.3$ and sizes $N=10^{5},10^{6},$ and $10^{7}$. We adopted $k_{c}=2\sqrt{N}$.}
	\label{fig:rho_near_threshold}
\end{figure}

We analyzed the epidemic prevalence for both discrete simulations and ELE mean-field theory. The infection rate was scaled according to the ELE prediction $\rho\sim \lambda{\Omega_1}-1$.
To analyze the critical exponent $\theta$, defined by $\rho\sim (\lambda-\lambda_c)^\theta$, of the discrete-time SIS dynamics and ELE, we consider the case in which $\gamma<2.5$, where the theory presents a great accuracy in predicting the epidemic threshold. We analyze the region in which $\lambda \Omega_{1}^{\text{ELE}}-1\leq 1$ and observed an almost perfect agreement between ELE and simulations, in which $\rho\sim (\lambda \Omega_{1}^{\text{ELE}}-1)^{\theta}$ with $\theta=1$, if $\lambda$ is not too close to the epidemic threshold, as shown in Fig.~\ref{fig:rho_near_threshold}. As previously reported in the continuous case~\cite{Diogo2020,Diogo2019}, the mean-field scaling with $\theta=1$ is confirmed if one is not too close to the epidemic threshold. The scaling $\theta=1$ remains valid for $g\in[0.1,1]$, while deviations are expected for $g\ll 1$ where the continuous-time limit occurs. The deviation of the linear scaling observed for finite $g$ (finite $\Delta t$), shrinks are the network size increases indicating that it is a finite-size effect due to falling onto the absorbing state. Here, a crucial difference between discrete- and continuous-time analyses is observed: the scaling region with exponent $\theta=1/(3-\gamma)$, rigorously obtained in continuous-time version~\cite{Mountford2013} and reproduced in extensive numerical continuous-simulations~\cite{Diogo2019}, is not observed in the discrete-time simulations discussed in Fig.~\ref{fig:rho_near_threshold}. This results represents an important issue when choosing the theoretical approach to model epidemic spreading on networks.

\begin{figure}[h]
	\centering
	\includegraphics[width=0.6\linewidth]{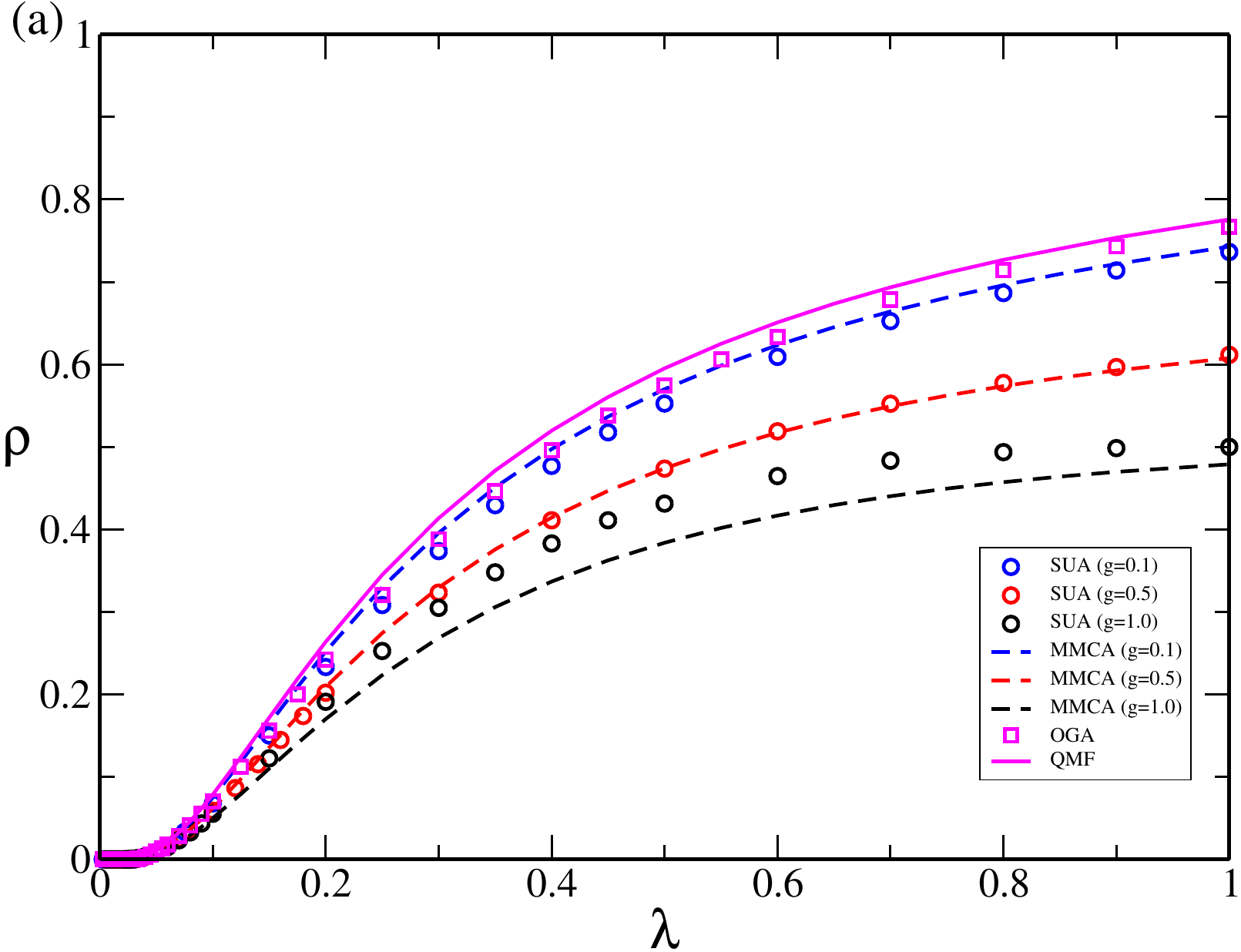}\\	
	\includegraphics[width=0.6\linewidth]{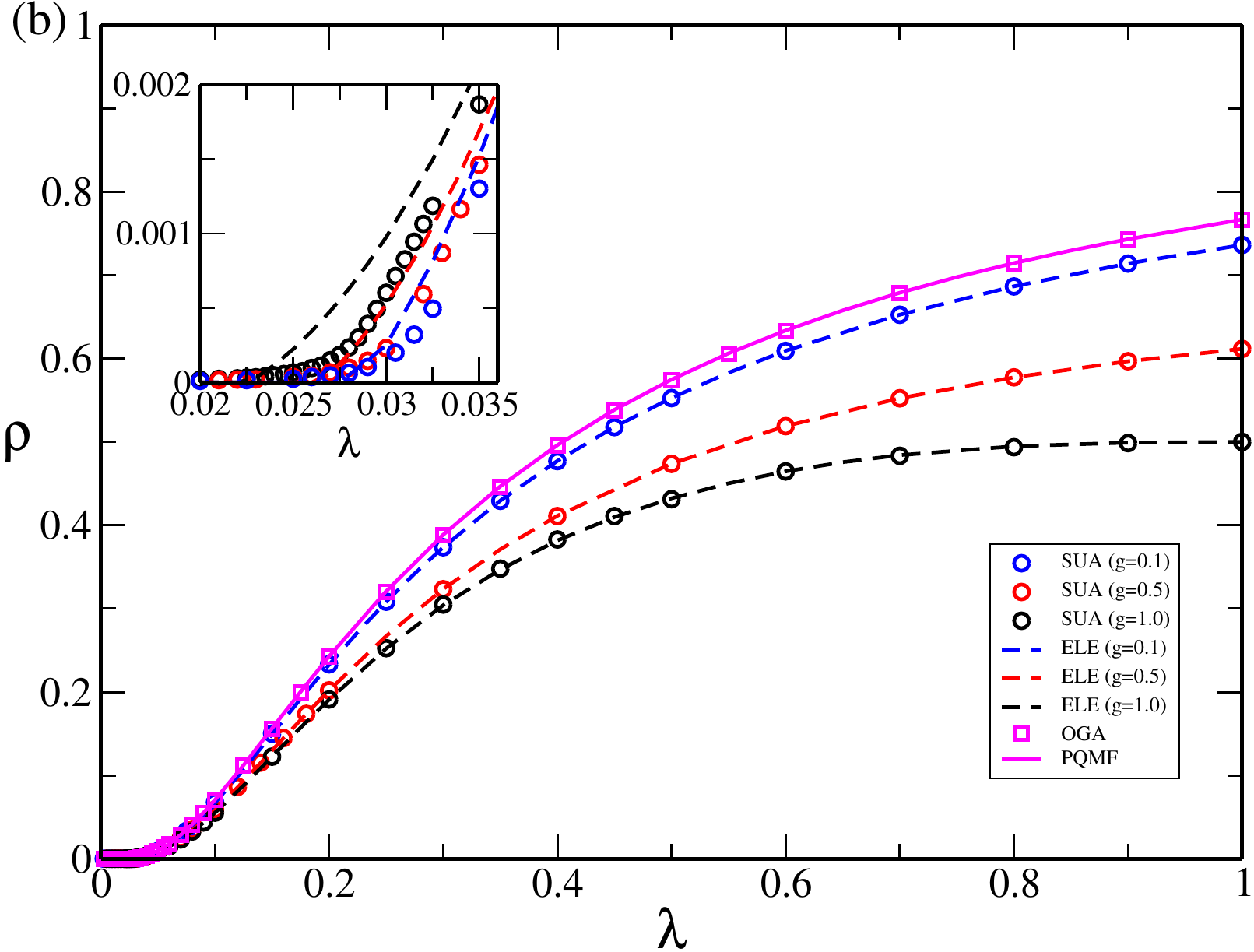}
	\caption{Prevalence as a function of the ratio $\lambda = \beta/\mu$ for discrete-time theory and simulations considering $\mu=1.0$ and uncorrelated power-law degree distribution network with $\gamma=2.8$. We adopted $ N=10^{6}$ and $k_{c}=2\sqrt{N}$. Both (a) MMCA and (b) ELE theories are presented. Inset shows a zoom in the low prevalence regime. Continuous-time limit of both theories (QMF and PQMF, respectively) are shown for sake of comparison. }
	\label{fig:high_prevalence}
\end{figure}
Considering the regime of high prevalence, we observe that MMCA and ELE present a dependence with $g$ and there is a convergence to  their respective theoretical continuous limit prevalence when $g\rightarrow 0$, as shown in Fig.~\ref{fig:high_prevalence}~(a) and Fig.~\ref{fig:high_prevalence}~(b), respectively.  Such as in the continuous-limit, the introduction of dynamical correlation leads to an almost perfect match between ELE and SUA simulations. For not too high prevalence, ELE theory performes better for smaller time steps, reaching the maximum performance in the continuous-time limit as shown in the inset of Fig.~\ref{fig:high_prevalence}~(b).  

\section{Conclusions}

The art of modeling epidemic processes on networks has been improved continuously with the implementation of elaborated aspects of the dynamical rules such as the precise role of absorbing states~\cite{Cota2017,Costa2021}, localization phenomena~\cite{Goltsev2012,Diogo2021} and non-Markovian nature~\cite{Feng2019,Boguna2014,vanMieghen2013a}. Mean-field approaches are primordial allies in the understating of epidemic processes on networks, and their accuracy must be probed using extensive and statistically exact stochastic simulations~\cite{Ferreira2012,Cota2017,Mata2013}. Computer simulations are implemented using discrete time: the continuous version considers asynchronous updates with variable time steps while the discrete one is performed with synchronous updated with fixed time steps. So, despite continuous-time approaches seeming more natural for actual epidemic processes, discrete-time versions are very popular~\cite{Gomez2010,Matamalas2018,Soriano2022} due to their easier computer implementation and also their flexibility to build models in terms of probabilities instead of rates. However, epidemic processes on networks can be very puzzling~\cite{Castellano2012,Mata2015} and dependent on the slightly different model details~\cite{Cota2018a}. Therefore, the role of synchronous and asynchronous approaches worthies investigation.

In the present work, we scrutinized the effects of discrete and continuous time versions of the SIS epidemic model considering stochastic simulations as well as one-node and pairwise mean-field theories.  Analyzing the epidemic threshold and the localization of the epidemic prevalence near the transition, we report that discrete-time approaches are dependent on the time-step size while differences between discrete and continuous cases disappear as the time-step goes to zero. While the previous finding is not surprising, we report a crucial difference between continuous and discrete time simulations: the epidemic prevalence near the epidemic threshold goes to zero following different scaling exponents $\rho\sim (\lambda-\lambda_\text{c})^\theta$ where the former agrees with the exact exponent $\theta=1/(3-\gamma)$ reported by Chatterjee and Durret~\cite{Chatterjee2009} while latter still provide the mean-field exponent $\theta=1$~\cite{Goltsev2012,Diogo2019,Matamalas2018}. Our results, therefore, raise important warnings on the choice of the theoretical and simulational approaches that can be present in other epidemic models or, more generally, dynamical processes on complex networks rather than the SIS model investigated in the present work.

\label{sec:conclu}

\begin{acknowledgments}

DHS thanks the support given by \textit{Funda\c{c}\~{a}o de Amparo \`{a} Pesquisa do Estado de S\~{a}o Paulo} (FAPESP)-Brazil (Grants No. 2021/00369-0 and No. 2013/07375- 0).  FAR acknowledges \textit{Conselho Nacional de Desenvolvimento Cient\'{i}fico e Tecnol\'{o}gico}  (CNPq)-Brazil (Grant 309266/2019-0) and FAPESP (Grant 19/23293-0) for the financial support given for this research. This research was conducted with the computational resources of the Center for Research in Mathematical Sciences Applied to Industry (CeMEAI) funded by FAPESP, Grant 2013/07375-0. SCF thanks the support by the CNPq (Grant 311183/2019-0) and \textit{Funda\c{c}\~{a}o de Amparo \`{a} Pesquisa do Estado de Minas Gerais} (FAPEMIG)-Brazil (Grant no. APQ-02393-18).

\end{acknowledgments}



\bibliographystyle{unsrt}
\bibliography{discrete_vs_continuous.bib}

\end{document}